\documentclass[11pt,a4paper]{article}

\usepackage{vipasustyle}
\usepackage{amsmath}
\usepackage{amssymb}
\usepackage{amsfonts}
\usepackage[normalem]{ulem}
\usepackage{color}
\usepackage{bbold}
\usepackage{ulem}
\usepackage{braket}
\usepackage{physics}
\usepackage{caption}

\title{Variational Preparation of the Sachdev-Ye-Kitaev Thermofield Double}
\author{Vincent Paul Su}

\date{\today}

\affiliation{Center for Theoretical Physics \& Department of Physics,

University of California, Berkeley, CA 94720, USA}

\emailAdd{vipasu@berkeley.edu}

\abstract{
  We provide an algorithm for preparing the thermofield double (TFD) state of the Sachdev-Ye-Kitaev model without the need for an auxiliary bath. Following previous work, the TFD can be cast as the approximate ground state of a Hamiltonian, $H_{\text{TFD}}$. Using variational quantum circuits, we propose and implement a gradient-based algorithm for learning parameters that find this ground state, an application of the variational quantum eigensolver. Concretely, we find quantum circuits that prepare the ground state of $H_{\text{TFD}}$ for the $q=4$ SYK model up to $N=12$.
}

\begin{document}

\maketitle

\section{Introduction}

The thermofield double (TFD) is a particularly special state in the AdS/CFT
correspondence \cite{maldacena97_large_n_limit_super_field_theor_super}.
Its duality to an eternal two-sided black hole was first discussed in Ref.~\citep{Maldacena_2003}.
This geometry features two copies of AdS that are connected by a
wormhole. One striking feature is the inability to causally signal from one CFT
to the other due to the spacelike nature of the wormhole.
However, in Ref.~\cite{Gao_2017}, the authors showed that a particular interaction of the two
CFTs rendered the bulk wormhole traversable, allowing causal probes to reach one
CFT from the other.

The Sachdev-Ye-Kitaev (SYK) model \cite{Sachdev_1993, kitaev2015} has been studied as a simple quantum mechanical model that exhibits conformal symmetry at low energies, with dynamics governed by a Schwarzian action \cite{Maldacena_SY_2017}.
In Ref.~\cite{maldacena2018eternal}, the authors explore the connection between SYK
and traversable wormholes in more detail.
One key result is that they provide a coupled Hamiltonian on the two systems whose ground state approximates the TFD.
In this work, we will consider the task of state preparation of the TFD of the
SYK model on Noisy Intermediate-Scale Quantum (NISQ) \citep{Preskill_2018} devices.

The general task of preparing the TFD of arbitrary theories was considered in Ref.~\cite{Cottrell_2019}. Again, the strategy was to construct a Hamiltonian whose ground states encoded the TFD structure.
Though the Hamiltonian in Ref.~\cite{maldacena2018eternal} can be thought of as a slightly
specialized version, we will use it in this work for its relative simplicity.
Both approaches consider the use of an auxiliary bath to adiabatically cool the system to its ground state.
Here, we instead take a variational approach, starting with a quantum circuit
ansatz whose parameters are tunable. This obviates the need for an auxiliary system.

A similar approach was used to construct the TFD of the Ising model
\cite{Wu_2019}. Shortly after, the TFD of the critical transverse field Ising
model (TFIM) on 6 qubits was implemented on ion traps \cite{zhu2019generation}.
Our approach will have two significant features for holography.  The first is
that larger qubit systems will eventually reach the semiclassical limit, as
opposed to the TFIM, whose central charge is small ($c=1/2$).  Second, our
variational approach does not rely on classical computers to perform
optimization of the quantum circuit, providing a path to a scalable procedure
for larger qubit systems.

In this paper, we will use variational quantum circuits to prepare the TFD of
the SYK model.
In Section~\ref{sec:TFD}, we review the physics of the thermofield double of the
SYK model as presented in Ref.~\cite{maldacena2018eternal}. The
TFD preparation approximately boils down to a ground state problem of a modified
Hamiltonian, with the approximation becoming exact in the large $N$ limit. In Section~\ref{sec:QC},
we provide background on variational circuits and how to optimize them.
In Section~\ref{sec:results}, we share the results of our variational procedure
applied to TFD preparation, making use of the excellent quantum simulation
package \texttt{Yao.jl}~\cite{luo19_yao}. We successfully prepare the approximate TFD on the
$N=8, q=4$ variant of SYK models, even in the presence of shot noise,
uncertainty that arises from estimating expectation values with a finite number
of sample. For $N=12, q=4$, we find a state whose energy is
within $\sim .2\%$ of the true ground state energy. For $N=16$, we were unable
to readily carry out our procedure due to classical computational resources.
We conclude with some remarks in Section~\ref{sec:discussion}.

\section{The Thermofield Double of SYK}\label{sec:TFD}

The thermofield double (TFD) at inverse temperature $\beta$ is defined for general quantum systems by entangling two copies of a quantum system in the following way.

\begin{equation}
  \label{eq:tfd}
  \ket{\text{TFD}(\beta)} = \frac{1}{Z^{1/2}_\beta}\sum_n e^{-\beta E_n/2} \ket{n}_L \ket{\bar{n}}_R
\end{equation}
where $n$ labels the energy eigenstates of a Hamiltonian that defines the system.
The subscripts $L, R$ correspond to the ``left'' and ``right'' systems from the
context of two sided black holes that appear in Ref.~\cite{Maldacena_2003}.
Upon tracing out either the left or right system, we are left with a thermal
state on the remaining system. The TFD can be thought of as a particular
purification of a thermal state on a single system.

In the following section, we will motivate the thermofield double of a particular quantum mechanical system, the SYK model.
In Section \ref{ssec:syk}, we discuss briefly the SYK model and some of the physics related to traversable wormholes.
In Section \ref{ssec:h_tfd}, we review recent literature on the preparation of the TFD, with specific application to the SYK model.

\subsection{SYK and Traversable Wormholes}\label{ssec:syk}

The SYK model at its core involves $N$ Majorana fermions with all-to-all $q$-body
interactions. Here we consider $q=4$ for concreteness. The Hamiltonian can
be written simply as follows

\begin{align}\label{eq:SYK}
  H_{\text{SYK}} = \sum_{ijkl} \mathcal{J}_{ijkl}  \psi_i \psi_j \psi_k \psi_l \, , \\
  \quad \mathcal{J}_{ijkl} \sim \mathcal{N}\left(  0, \frac{12  J^2  } {  N^{3}}\right) \, ,
  \end{align}
where the strength of the couplings of $\mathcal{J}_{ijkl}$ are {i.i.d.\
  Gaussian}. They are distributed normally about 0 with a width set by a parameter $J$.

The SYK model was first proposed as a quantum model of holography by Kitaev
\cite{kitaev2015} with dynamics that had an analytic solution in the large $N$
and large $\lambda \equiv \beta J$ limit. More recently, its connection to gravity
has been studied in the duality to nearly AdS$_2$ spacetimes.
In Ref.~\cite{maldacena2018eternal}, the low energy theory of two copies of the SYK model
coupled by an interaction is compared against the low energy theory of nearly
AdS$_2$ solutions that describe traversable wormholes discussed in Refs.~\cite{Gao_2017, Maldacena_SY_2017}. In the large $N, \lambda$ limit, they show the action is the same, indicating
the same gravitational physics at play. See
Refs.~\cite{maldacena16_confor_symmet_its_break_two,jensen16_chaos_ads_holog,kitaev17_soft_mode_sachd_ye_kitaev,cotler16_black_holes_random_matric} for more on the
connection between SYK and gravity.

This duality is quite remarkable because it allows us to probe our understanding
of quantum gravity by controlling simple quantum mechanical systems with precision.
In Ref.~\cite{gao2019traversable}, the authors leverage this connection by proposing a traversable wormhole experiment using SYK operators.
The extent to which this experiment on a quantum device could probe our
geometric interpretation of the dual spacetime is still an open question.
However, if one could explicitly quantify how to probe the interior wormhole geometry and implement this on a
quantum device, it would be an amazing
experimental check on our understanding of quantum gravity, not to mention much
more feasible than collapsing a black hole in the lab.
As a starting point, however, they require the TFD state, whose preparation is the aim of this work.

\subsection{The TFD as a Ground State}\label{ssec:h_tfd}
Given the special role that the TFD plays in quantum systems, one very natural
question that arises is how to actually prepare such a state. The general task was
considered in Ref.~\cite{Cottrell_2019} and approximately for the SYK system in
Ref.~\cite{maldacena2018eternal} (see also \cite{maldacena19_syk_wormh_format_real_time}).
In short, one takes the Hamiltonians of the left and right
systems with an interaction term to favor the correct entanglement pattern. In
the light of optimization, one may view this interaction as a Lagrange multiplier.
Both references propose preparing the TFD by using an external bath to cool the system to the
ground state of this modified Hamiltonian.

In Ref.~\cite{maldacena2018eternal}, the authors show that the TFD is approximately given by the ground state of the following Hamiltonian
\begin{align}
  \label{eq:tfd_ham}
  H_{\text{TFD}} &= H_{L, \text{SYK}} +  H_{R, \text{SYK}} + H_{\text{int}} \, , \\
  H_{\text{int}} &= i\mu \sum_j \psi_L^j \psi_R^j \, ,
\end{align}
where the factor of $i$ ensures the hermiticity. The strength of this interaction $\mu$ plays two roles. One can consider $\mu$
as describing a relevant deformation away from the conformal limit of the SYK
model. Secondly, it sets the scale of the inverse temperature $\beta$.

The sense in which the TFD is approximately the ground state of
$H_{\text{TFD}}$ is characterized by the overlap, or fidelity. This overlap
approaches 1 in the classical limit,
\begin{equation}
  \label{eq:parameter_regime}
  1 \ll \lambda , N \, .
\end{equation}
The interaction strength $\mu$ should also be small relative to the energy scale
$J$ for finite $q$.

Away from the large $N$ limit, the leading correction to the overlap is
calculated in Ref.~\cite{maldacena2018eternal}. In
Ref.~\cite{maldacena19_syk_wormh_format_real_time}, they distinguish between the
TFD of the two decoupled systems and the ground state of $H_{\text{TFD}}$ by
referring to the latter as the ``SYK wormhole''. In this work, we make the
simplification that these are the same, i.e. that we are in a classical regime.
Although current systems are far from having a large $N$ number of qubits, our
procedure could be repeated as the quantum hardware advances.
We will leave investigation of quantum corrections to future direction and work under
the assumption that we wish to prepare the ground state of $H_\text{TFD}$.

To summarize this section, we have reduced the problem of preparing the SYK
thermofield double to a problem of finding the ground state of a particular
Hamiltonian, $H_\text{TFD}$. In contrast to previous work that suggest the
use of an external bath to cool the system, we will take a variational approach
without any additional degrees of freedom.

\section{Variational Quantum Computing}\label{sec:QC}

The aim of this section is to introduce the framework of variational quantum computing.
We will work with the circuit model of quantum computation; for a review, see Ref.~\cite{NielsenChuang}.
The core algorithm we will leverage is known as the variational quantum
eigensolver (VQE), originally
applied to quantum chemistry \cite{peruzzo13_variat_eigen_solver_quant_proces}. However, the task of finding the
minimal eigenvalue of a Hamiltonian has myriad applications.

From a high-level perspective, variational methods may be better thought of in terms of
optimization, which is slightly different from the traditional view of
algorithms that have a fixed procedure. Whereas a typical algorithm might
generate a ground state by a fixed computation, variational methods seek to
optimize parameters that minimize a corresponding objective function. For us,
the Hamiltonian serves as the objective function.

In the context of NISQ Devices, variational
algorithms are useful for several reasons reasons. First, quantum algorithm
design is extremely difficult. Piecing together quantum gates to make a meaningful
computation is as unintuitive as writing a calculator with access to only AND
and OR gates.

Second, circuit depth is a critical limiting factor in the NISQ era.
Gate compilation, or translation between universal gate sets, leads to prohibitively large
overhead in terms of gate count. Variational algorithms with
gates that are native to the architecture keep the circuit depth minimal.

A third benefit of variational algorithms is that they are more error resilient
since hardware teams are optimizing the gates they advertise. Additionally,
while universal gate sets are equivalent, there is still generically an $\epsilon$ error when
approximating a gate using a finite number of gates from a different gate set.

In Section~\ref{ssec:PQC}, we describe \textit{parameterized quantum circuits} (PQC),
circuits whose sequence of gates are fixed but have tunable
parameters, such as angles of a Pauli rotation. We will introduce a particularly
simple choice of PQC.
In Section~\ref{ssec:gradients}, we discuss how to optimize such parameterized
quantum circuits with the help of classical computation. However, as we
explain in Section~\ref{ssec:param_shift}, the parameter shift rule will allow us to significantly
downplay the role of classical optimization.

\subsection{Parameterized Quantum Circuits}\label{ssec:PQC}

\begin{figure}
    \centering
    \includegraphics{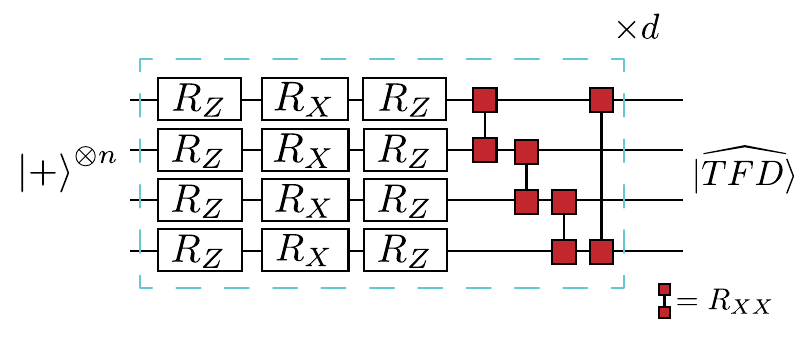}
    \caption{A simple parameterized quantum circuit (PQC) on $n=4$ qubits composed of
      single qubit rotations and nearest neighbor two qubit gates. Each of the
      rotations is described by a continuous parameter $\theta$. On the
      diagram, $R_X = \exp(\frac{i}{2}\theta X)$. Parameters are not necessarily
      shared between gates. This sequence of gates is repeated $d$ times, indicating the depth of the
      circuit. The PQC acts on a fixed input state and produces our estimate of
      the TFD state. We can thus switch between viewing the variational object
      as either the circuit $U(\boldsymbol{\theta})$ or the state as in
      Eq.~\eqref {eq:psi_variational}. This is the form of ansatz we consider in
    the Section~\ref{sec:results}. One can also replace the Ising XX
    ($R_{XX}$) gate with other 2 qubit entanglers and get similar results.}
    \label{fig:pqc}
\end{figure}

In short, parameterized quantum circuits are a variational ansatz for a quantum circuit composed of several gates. Some of the gates in our ansatz will have
parameters which determine the amount of rotation for a single qubit gate.
Our criteria for success will amount to finding parameters that
minimize the energy of some Hamiltonian $H$. For a fixed input state, one may then view this circuit
ansatz as a wavefunction ansatz.

Here we describe a rather simple ansatz composed of single qubit rotation
gates and CNOT gates. The PQC we will consider in Section~\ref{sec:results} is outlined in
Figure~\ref{fig:pqc}. Let us consider a system with $N$ qubits. One layer of gate
applications will be composed of parameterized single qubit rotations on each
qubit followed by a cyclic application of CNOTs, which are not parameterized.
Our ansatz will then be $d$ such layers.

Phrased in this way, both our circuit and, consequently, the resulting state
are parameterized by the angles $\boldsymbol{\theta}$
\begin{equation}
  \label{eq:psi_variational}
 \ket{\psi(\boldsymbol{\theta})} = U(\boldsymbol{\theta}) \ket{0} \, ,
\end{equation}

where $U(\boldsymbol{\theta})\ket{0}$ describes our circuit acting on a fixed input
state in the computational basis. The energy would then simply be the
expectation of the hamiltonian in this state.

\begin{equation}
  \label{eq:E_variational}
 \widehat{E}(\boldsymbol{\theta})  = \bra{\psi(\boldsymbol{\theta})} H \ket{\psi(\boldsymbol{\theta})} \, .
\end{equation}

In the next section we will describe the procedure for how to optimize the
parameters with respect to the energy.

\subsection{Gradient Descent Review}\label{ssec:gradients}

From an optimization view, we are faced with the task of minimizing a scalar function $\widehat{E}$ of multiple parameters.
The first naive algorithm to perform this is gradient descent, where one updates
the parameters based on the gradient at some initial point.
If $\widehat{E}$ is convex, then vanishing gradient
guarantees that you have found a global minimum. In the non-convex case,
vanishing gradient implies you are in a local minimum, but there are no
guarantees on optimality.

Algorithmically, there is only a single computation
involved, which is the gradient update rule
\begin{align}
  \label{eq:gradient_descent}
  \boldsymbol{\theta}_{t+1} &= \boldsymbol{\theta}_{t} - \eta \Delta \boldsymbol{\theta} \, , \\
  \Delta \boldsymbol{\theta} &= \frac{\partial \widehat{E}(\boldsymbol{\theta})}{\partial \boldsymbol{\theta}} \, .
\end{align}

Here $\eta$ is a parameter often referred to as \textit{step size} which can
also vary in time. Many improvements to the basic gradient descent rule have
been discussed in the machine learning literature, with many ideas stemming from
physics. See Ref.~\cite{ruder16_overv_gradien_descen_optim_algor} for a concise overview. We will
simplify the discussion by first considering the basic form of gradient descent,
though we will use a slightly more sophisticated version in Section~\ref{sec:results}.

Let us first consider some basic computational aspects of the gradient of a
circuit.\footnote{Precisely (in the Schrodinger picture) what me mean is the gradient of an observable in the
state prepared by the circuit. In the Heisenberg picture, we are measuring an
observable that is a function of the circuit parameters.}
The naive matrix representation of the circuit involves $2^N \times 2^N$ for a system of $N$ qubits.
Classically, one could search over parameter space to find the optimal angles, removing the need for gradient descent if the number of qubits is not too large.
However, this very quickly becomes intractable to even store in memory.

By contrast, the gradient storage depends only on the number of parameterized gates.
In the simple architecture discussed in the last section this would be $O(Nd)$ entries.
This back of the envelope estimation is just to emphasize that relying on
classical computers to do the optimization of parameters in an exponentially
large space is difficult without clever manipulation. On the other hand, the
gradient information can be stored and updated easily, provided the gradient can
be estimated efficiently.

\begin{figure}
    \centering
    \includegraphics[width=.8\textwidth]{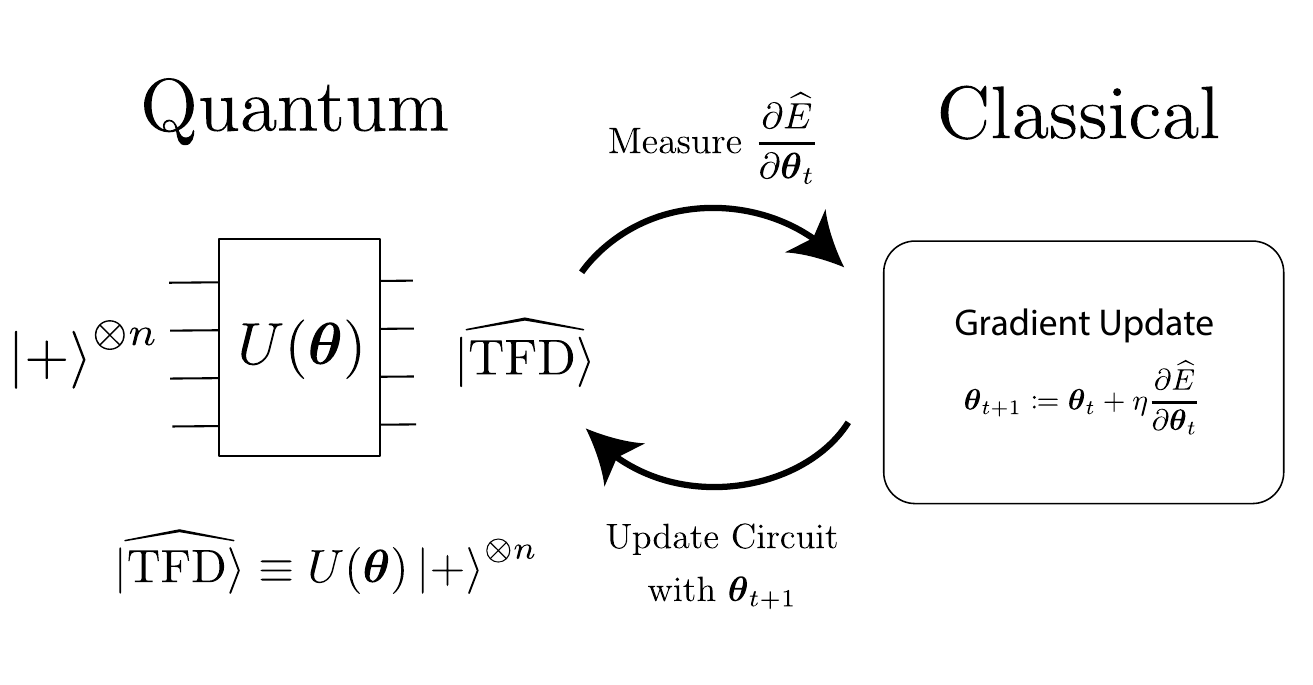}
    \caption{A schematic representation of hybrid quantum classical computation.
      Here we would like to learn the best parameters for our quantum circuit.
      Measurements from the circuit inform how
      the classical computer produces new parameters to modify the quantum
      circuit. An
      extremely important aspect of our procedure is that the gradient
      information is obtained directly from measurements of the quantum circuit,
      circumventing the need for calculating gradients of matrices of
      exponential size on a classical computer.}
    \label{fig:hybrid}
\end{figure}


It turns out that indeed one can get estimates of the gradient directly from the
quantum hardware, and importantly, from the same architecture as the
ansatz. Thus, our scheme will be a hybrid one, using the parameterized quantum
circuit to encode our state and provide gradient information, with a classical
computer as an aide to update the circuit parameters given gradient information.
This is summarized in Figure~\ref{fig:hybrid}.

We briefly discuss two approaches for obtaining the gradient from the hardware, one referred to as \textit{finite
  differences} and the other called the \textit{parameter shift rule}.
The finite difference method uses the limit definition of derivatives by testing the response to a perturbation in each of the parameters.
This works best if the perturbation is infinitesimal.
In the context of NISQ devices, this perturbation may easily fall under the noise threshold, rendering the measurements meaningless.

The parameter shift rule, on the other hand, involves large, finite variations
(e.g. $\pm \pi/2$). Under the appropriate choice of gates, this gives an
unbiased estimate of the gradient. In other words, the expectation value matches
the true analytic gradient. We elaborate more on the parameter shift rule in the next section.

\subsection{Parameter Shift Rule }\label{ssec:param_shift}

The parameter shift rule allows us to compute the gradient of the expectation value of an observable with
respect to the parameters of a PQC. The technique is quite remarkable and worth
highlighting. First, it is not obvious at all that the output of \textit{any}
quantum circuit will give information about how to tune parameters of a
variational circuit. Second, of great practical importance, is the amazing
fact that it is the \textit{same} variational circuit that gives gradient
information about the quantum circuit! Both of these are consequences of the
parameter shift rule. We start by discussing a toy example, following the original work of
\cite{Mitarai_2018}\footnote{See also \url{https://pennylane.ai/qml/glossary/parameter_shift.html}}.
Later we address questions of generalizing to the relevant setup.

For the sake of simplicity, let us consider a single qubit which can be visualized on the Bloch sphere.
Our parameterized quantum circuit will consist of a single Pauli rotation gate
\begin{align}
  R_X(\theta) \equiv \exp(iX\theta/2 ) \, ,
\end{align}
where we use interchangeably the notation $X$ for the Pauli operator $\sigma^x$.

Suppose we are interested in minimizing the expectation of the observable $Z$. Clearly, the optimal qubit state which minimizes $\ev{Z}$ is the $\ket{1}$ state at the south pole of the Bloch sphere. In the language described in previous sections, our circuit is $U(\theta) = R_X(\theta)$ and produces the variational state $\ket{\psi} = U(\theta)\ket{0}$.

To minimize the expectation of $Z$, we would like to calculate the gradient with respect to $\theta$.

\begin{equation}
    \pdv{\ev{Z(\theta)}}{\theta} = \pdv{\ev{U^\dag(\theta) Z U(\theta)}{0}}{ \theta} \, .
\end{equation}
By the product rule,
\begin{equation}
    \label{eq:product_rule}
    \pdv{\ev{U^\dag(\theta) Z U(\theta)}{0}}{ \theta} = \frac{i}{2}\ev{U^\dag(\theta)(XZ - ZX)U(\theta) }{0} \, .
\end{equation}
Using the Pauli algebra, one can express the commutator $[X,Z]$
\begin{equation}
    [X, Z] = i \left(U^\dag\left(\frac{\pi}{2}\right)Z U\left(\frac{\pi}{2}\right) - U^\dag\left(-\frac{\pi}{2}\right)Z U\left(-\frac{\pi}{2}\right) \right)
\end{equation}
Plugging back into Eq~\eqref{eq:product_rule},
\begin{equation}
    \pdv{\ev{Z(\theta)}}{\theta} = \frac{1}{2} \left(\ev{U^\dag(\theta) U^\dag\left(\frac{\pi}{2}\right)Z U\left(\frac{\pi}{2}\right) U(\theta)}{0}  - \ev{U^\dag(\theta) U^\dag\left(-\frac{\pi}{2}\right)Z U\left(-\frac{\pi}{2}\right) U(\theta)}{0} \right)
\end{equation}
Note that the extra applications of $U$ can be absorbed into the original angle, leaving us with a simple form.
\begin{equation}
\pdv{\ev{Z(\theta)}}{\theta} = \frac{1}{2}\left( \ev{Z(\theta^+) } - \ev{Z(\theta^-}\right)
\end{equation}
where $\theta^\pm = \theta \pm \pi/2$. We can see now that the derivative is given by the difference of two expectation values using exactly the same form of the quantum circuit, as promised at the beginning of the section.

This toy example can easily be extended to more complicated circuits and observables. First let us address the
issue of multiple angles and multiple gates. Decomposing the full unitary into a
product of unitaries, the partial derivative only acts on terms which contain
the angle in question. All other gate applications in the sequence can therefore
be absorbed into either the state or the measurement. One then repeats this for
all parameters in the circuit to obtain all gradient elements.

Second, we are
ultimately interested in minimizing a Hamiltonian that contains many terms. This is also not an issue by the linearity of expectations and derivatives. Namely,

\begin{equation}
    \pdv{\ev{H}}{\boldsymbol{\theta}} = \sum_i \pdv{\ev{P_i}}{\boldsymbol{\theta}} \, ,
\end{equation}
where we have used the fact that Pauli strings form a complete basis for
operators. We can then apply the parameter shift rule to each term in the sum.

It may seem overly restrictive to consider gates which are generated by the
Pauli matrices. However, this was generalized in
in Ref.~\cite{Schuld_2019}, where they consider the primary task of evaluating
these circuit gradients on actual hardware. They first generalize the parameter
shift rule beyond tunable gates which are generated by the Pauli's. It suffices
to consider gates whose generators have up to two eigenvalues. As they point out,
three of Google's native Xmon gates satisfy this criteria. Alternatively, for
circuit architectures which do not fit this structure, one can apply
\textit{linear combination of
  unitaries}~\cite{childs12_hamil_simul_using_linear_combin_unitar_operat} using
an ancilla qubit to get estimates of the gradient.

In concert with the quantum circuit, a classical computer is still required to
carry out the gradient update step. To our knowledge, there are no quantum
algorithms for numerically adding vectors that a person with finite resources would choose over a
classical computer.
Thus, the full procedure requires a \textit{hybrid} approach to computation,
summarized in Figure~\ref{fig:hybrid}.

Notably, previous discussions of this
classical-quantum hybrid model have involved using the classical computer for
optimization. By using the parameter shift rule to obtain gradients, we have
downplayed the role of classical computers in this hybrid model. This is
critical for applying this algorithm to larger qubit systems.

\section{Preparing the TFD on a Near Term Quantum Device}\label{sec:results}

In this section, we apply our variational algorithm to find the ground state
of $H_{\text{TFD}}$ as in Eq.~\eqref{eq:tfd_ham}.
First in Section~\ref{ssec:JW}, we briefly overview the Jordan Wigner
transformation that maps fermionic degrees of freedom to qubits.
Once the Hamiltonian is expressed in terms of qubits, we can use the PQC described in Figure~\ref{fig:pqc}.
Subsequently in Section~\ref{ssec:sims}, we outline the numerical simulation
procedure. In Section~\ref{ssec:results}, we
present the main result of the paper, which is finding the
ground state of the $H_\text{TFD}$ for $N=8$.
Excitingly, as few as $d=2$ layers can achieve the ground state for some
instantiations of the SYK model. For $N=12$, our variational approach rapidly
approaches the near degeneracy at the ground state energy. Finally, to bring this even one step closer to
physical implementation, in Section~\ref{ssec:shots} we include the effects of \textit{shot noise} in our
simulation.

\subsection{Jordan-Wigner Transformation}\label{ssec:JW}
By far, qubits are the dominant computing paradigm in NISQ devices. In order to
run our algorithm, we will need to express our Hamiltonian in terms of qubits
rather than Majorana fermions. We review the Jordan-Wigner
transformation that precisely enables this.\footnote{See Ref.~\cite{Nielsen_2005} for an excellent review.}

The Jordan-Wigner transformation is a duality mapping between spinless fermionic
operators to non-local Pauli operators. For Majorana fermions, which
are their own antiparticles, this mapping can be written as
\begin{align}
  \label{eq:jw}
  \psi_l = \left( \prod_{j=1}^{\tilde{l}-1}\sigma_{j}^z \right) \sigma_{\tilde{l}}^{\alpha_l} \,
\end{align}
where $\tilde{l} = \lfloor{(l+1)/2}\rfloor$ and $\alpha_l$ denotes $x$ if $l$ is even
and $y$ otherwise. In short, each of the Majorana operators gets mapped to a
Pauli string.

Products appearing in the SYK Hamiltonian likewise get
mapped to products of these Pauli strings. Notably, there are many such
redefinitions that preserve the algebra structure of the operators. For example,
multiplying the definition by $\prod_{i=1}^{N}\sigma_i^z$ preserves the
algebraic structure. There are other ways to encode the operators, such as the
Bravyi-Kitaev~\cite{Bravyi_2002} transformation which has been shown to have typically
shorter length Pauli strings, but whose mapping is less intuitive.

Equipped with this redefinition, we can express $H_\text{TFD}$ in terms of
qubits, which we temporarily refer to as $H_\text{qubit}$.\footnote{The OpenFermion package
  \cite{mcclean2017openfermion} was used to automate this conversion.} Since $H_\text{TFD}$ is a sum
over products of Majorana operators, $H_\text{qubit}$ will be a sum over Pauli
operators, which are amenable to the parameter shift rule alluded to in the
previous section. Henceforth we will drop the distinction between $H_\text{TFD}$
and $H_\text{qubit}$.

The number of qubits required is $N$. In terms of our original system, $N$ was
the number of Majorana fermions on a single copy of the theory. In the
thermofield double, defined on two copies of the theory, we have a total of $2N$
Majoranas. However, the Hilbert space dimension of each Majorana fermion is
$2^{1/2}$. For a total of $2N$ Majorana fermions, this is equivalent to $N$
qubits. Thus the parameter $N$ denotes the number of Majoranas on a single
``side'' of the TFD and the number of qubits in our discussion.

\subsection{Quantum Circuit Simulation Overview}\label{ssec:sims}
In this section, we provide an overview of the full procedure of our protocol.
For more
details on the exact numerical procedure, we make the code publicly available.\footnote{\url{https://github.com/vipasu/SYK-TFD}}
Simulations were carried out via the quantum simulation package \texttt{Yao.jl}~\cite{luo19_yao}.

At a high level, we sample the couplings $\mathcal{J}_{ijkl}$ that define our instance of the SYK
model. This Hamiltonian is then shared for two (L/R) systems and coupled by an
interaction term of the form specified in Eq.~\eqref{eq:tfd_ham}, yielding a
combined Hamiltonian $H_{\text{TFD}}$ whose ground state we are interested in
finding. The Hamiltonian is mapped to a sum of Pauli strings. Our ansatz for
the state is specified by a choice of PQC described in Figure~\ref{fig:pqc}, initializing all parameters to 0. We
then update our ansatz by taking gradients of $H_{\text{TFD}}$ with respect to
the circuit parameters. Though we described the vanilla flavor of gradient
descent in Section~\ref{ssec:gradients}, numerically we used a variant known as
Adam \cite{kingma14_adam} that takes into account the gradient information of
previous iterations.\footnote{We made use of the \texttt{Flux.jl} package \cite{innes:2018} for
  implementation of Adam.}

\begin{figure}
  \centering
  \includegraphics[width=.85\textwidth]{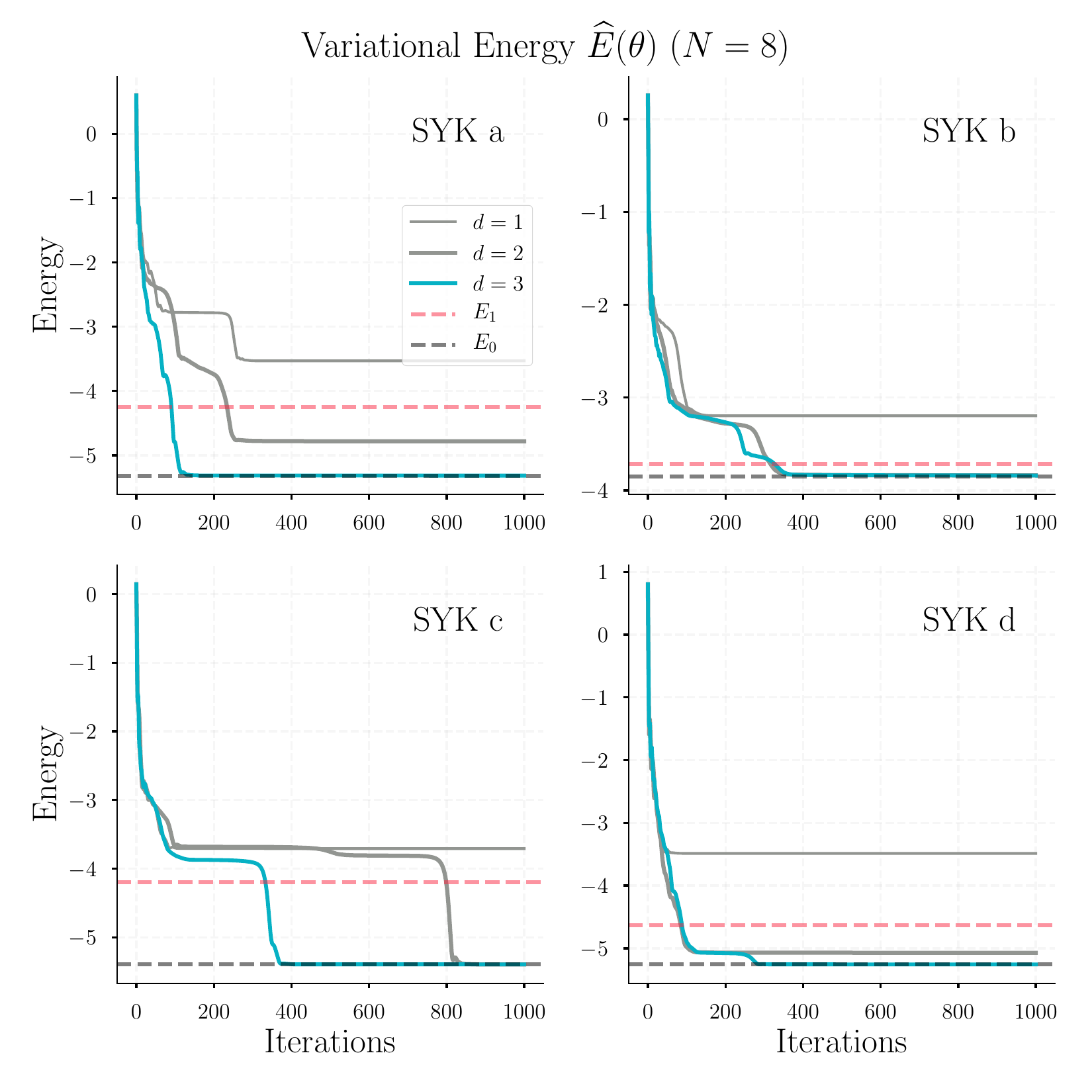}
  \caption{We plot the variational energy achieved by our parameterized
    circuit for four instantiations of the SYK model on $N=8$ qubits. For
    each of the Hamiltonians, we plot the energies of the ground state and
    first excited state. We show the variational energy for our PQC ansatz
    with different depths. As few as 2 layers can reach a
    variational energy below the first excited state in some cases. We could
    generically achieve the ground state with 3 layers of our PQC.}
  \label{fig:training}
\end{figure}

We again stress that the gradient information can be obtained from the quantum
circuit itself.
This removes the need to do any classical computation in large
hilbert space dimension. The classical part simply keeps track of the parameters
of the quantum circuit and updates them based on gradient information provided
by the quantum circuit.
In this work, we assumed access to the analytic gradient (e.g. perfect
knowledge). In practice, this will be thwarted both by the \textit{shot noise} of
calculating expectation values with a finite number of samples, and also by
errors in the gate applications. We provide the results of
repeating the experiments in the presence of shot noise in Section.~\ref{ssec:shots}.

Finally, for small enough system sizes (up to 12 qubits), we compare the
variational energy achieved with the spectrum obtained by exact diagonalization
of $H_{\text{TFD}}$.




\begin{figure}
  \centering
  \includegraphics[width=.85\textwidth]{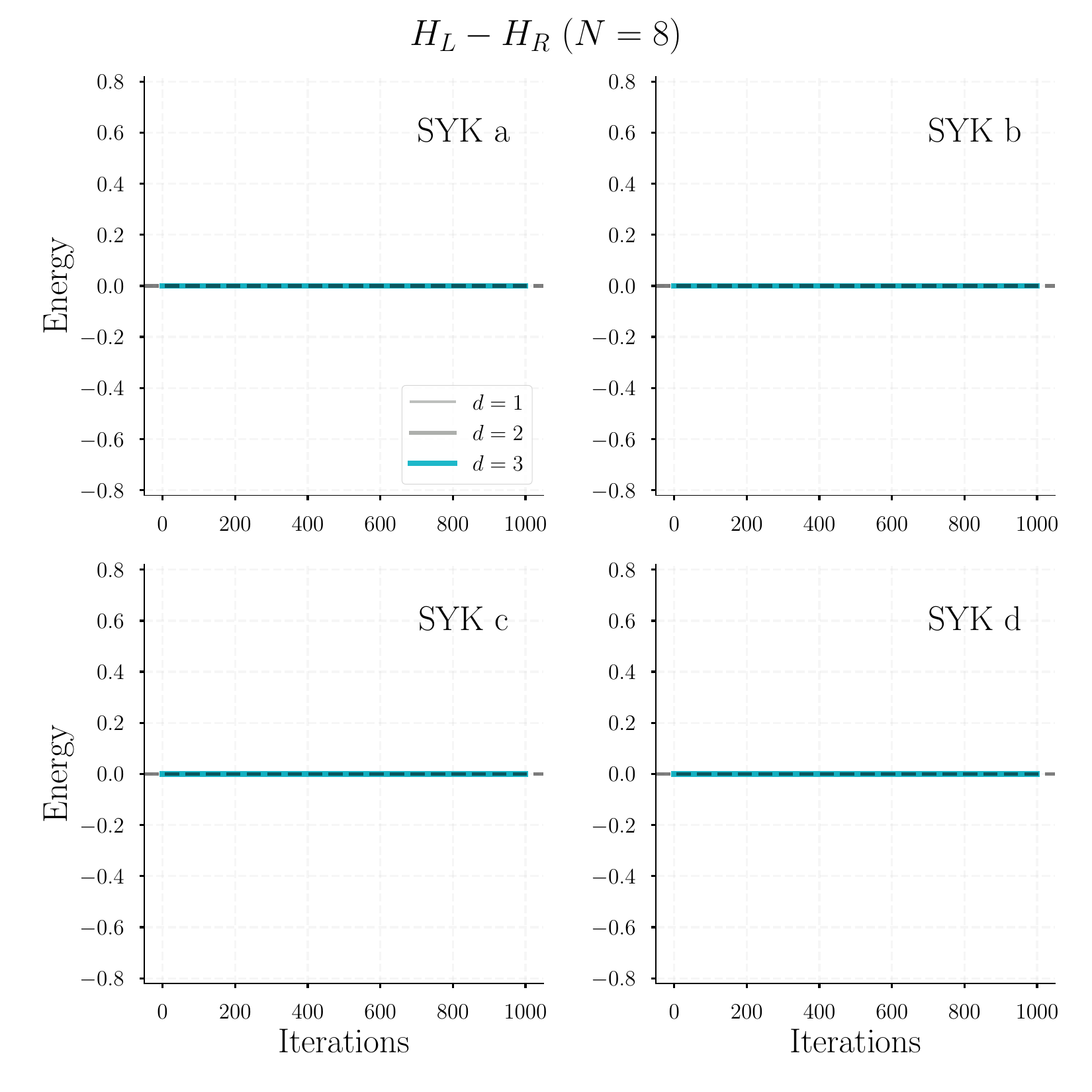}
  \caption{A parity diagnostic of the TFD state. The sum $H_L - H_R$ should
    have expectation 0 for the true TFD. This is true for any parity invariant
    state, so it is necessary but not sufficient. The Ising gate seems special
    in that other choices of entangling 2 qubit gates, didn't necessarily find
    a parity invariant state. In the presence of noise, the circuit starts to
    violate parity as well (as in Figure~\ref{fig:annihilator_shots}). In this work, we are simply using it
    as a diagnostic, but one may consider adding objective terms that try to
    minimize the square of this in addition to minimizing the energy of $H_{\text{TFD}}$. }
  \label{fig:annihilator}
\end{figure}
\subsection{Results with Analytic Gradients}\label{ssec:results}

In this section we share the findings of our numerical procedure.
We successfully
find the ground state of $H_\text{qubit}$ for the $N=8, q=4$ SYK model.
As an additional benefit of this variational computation, we have an explicit circuit
that can be implemented on hardware.

To verify that our variational state is indeed the ground state, we plot the
energy compared to the ground state energy obtained by exact diagonalization. As
an additional check, we track the expectation of $H_L - H_R$ in our state.
For system sizes of $N=8$, we can show our variational ground state gets
well below the first excited state energy, indicating that there is high, if not
perfect, with the ground state. We will come back to the case of $N=12$ momentarily.

\begin{figure}
  \centering
  \includegraphics[width=.85\textwidth]{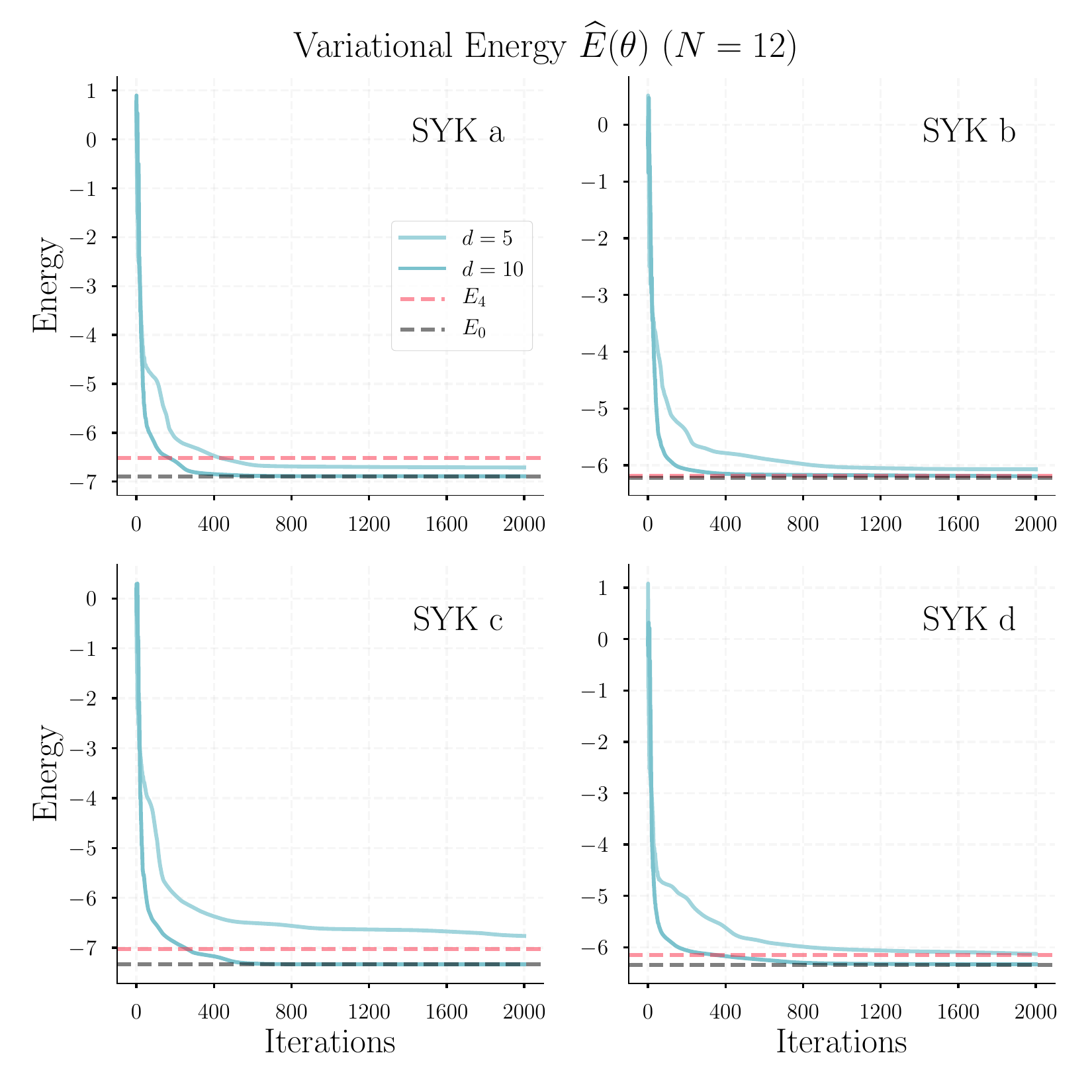}
  \caption{Variational Energy achieved on $N=12$.
    There is a near degeneracy in the spectrum, which is shown in
    Figure~\ref{fig:spectra}. The depth required was not significantly more. For our ansatz, a PQC of depth $d$ reaches a
    strict subset of states compared to depth $d+1$ since the last layer of
    rotations could have 0 angle. However, higher depth circuits did not always have lower variational energy at every training iteration, hinting
    to the subtleties in navigating the optimization landscape.}
  \label{fig:training12}
\end{figure}
\begin{figure}
  \centering
  \includegraphics[width=.85\textwidth]{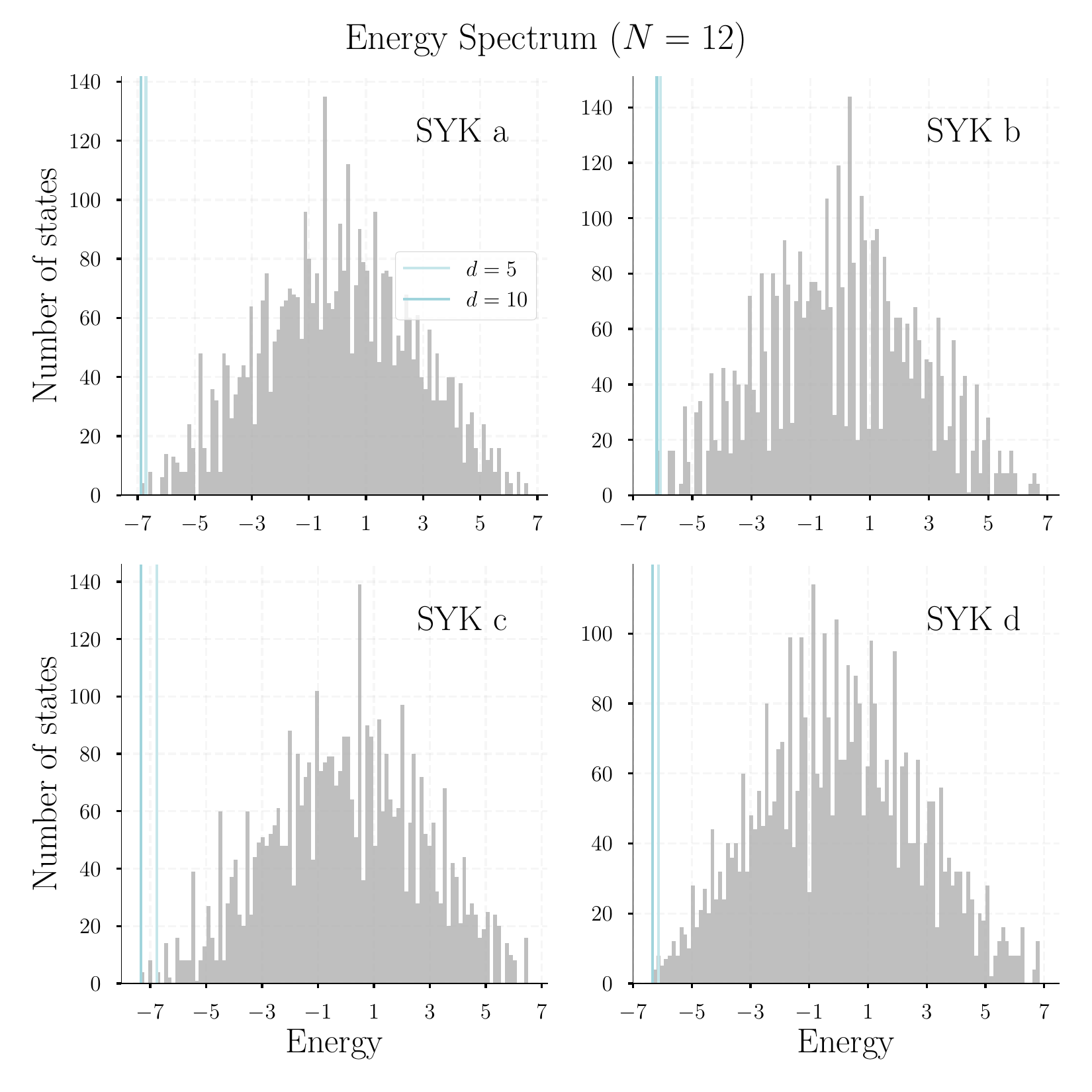}
  \caption{Spectrum obtained by exact diagonalization in a Hilbert space of
    dimension $2^{12}$. Vertical lines denote the
    minimum energy achieved by PQCs of varying depth. Interestingly, there is a
    degeneracy of states near the ground state energy. This is an indication
    that the more general form of $H_{\text{TFD}}$ in Ref.~\cite{Cottrell_2019}
    could be more apt.
  }
  \label{fig:spectra}
\end{figure}
For $N=8$, fewer than 500 gradient step
updates with depth $d=3$ worked well enough to produce the ground state of
several instantiations of the SYK Hamiltonian, as verified in
Figure~\ref{fig:training}. As few as $d=2$ layers was enough to reach energies below the first excited state.


An additional verification that the ground state has been achieved is by
checking the quantity $H_L - H_R$. One can straightforwardly verify that this
should annihilate the TFD state as in Eq.~\eqref{eq:tfd}. Indeed, in Figure~\ref{fig:annihilator}, we
track this quantity as a function of gradient iteration.
Our choice of the Ising entangling gate was partially motivated because it
maintained parity throughout the whole training. When repeated for different
choice of two qubit gates, this quantity fluctuated much more.

Though it was not necessary for this
example, one could in principle use the square of this quantity as a Lagrange multiplier in
the energy minimization. Since $(H_L-H_R)^2$ would also be a sum of Pauli strings,
this would likewise be amenable to the gradient techniques described in Section~\ref{ssec:param_shift}.


For $N=12$ qubits, we can run the same procedure, simply adding more terms to
the Hamiltonian and expanding our PQC to include more qubits. For sufficiently
large depth $d$, our procedure finds a state whose energy matches the ground
state energy to within $\sim .2 \%$. These results are summarized
in Figures~\ref{fig:training12} and \ref{fig:spectra}.


To recap, we have shown that a simple quantum circuit with reasonable depth
can prepare the approximate TFD for the SYK model on $N=8$ qubits. For $N=12$
qubits, we can numerically simulate a quantum circuit that approaches the ground
state energy, though more work would be required to bring it to a form
implementable on a NISQ device. Larger system sizes were not readily comparable via
exact diagonalization, so one would likely need to find a way to verify the
variational state, for example by the parity check and possibly by matching
thermal correlators.

For larger system sizes, we again stress that the variational procedure can
still implemented on quantum devices. This is in line with the ultimate goal of finding ways to use quantum
computers to probe physics we cannot already access on a classical computer. This
hinged on the fact that all of the
gradient information came from the quantum circuit itself. Thus, if the quantum
hardware can carry out the ansatz circuit, it can also provide the gradient
information necessary to perform this procedure.

\subsection{Learning in the Presence of Shot Noise}\label{ssec:shots}
At the core of this algorithm is the fact that gradients can be obtained from
the quantum circuit by measuring expectation values. Within a simulation, the
expectation values can be computed exactly. On a physical device, however, one must take into
account the noise introduced by having to average a finite number of
measurements, or \textit{shots}. This effect is called shot noise. In this section, we
will demonstrate that our algorithm can find the ground state in the presence of
this noise.

Let us precisely distinguish the noise we account for here. On a quantum
device, the expectation value of an observable requires several measurements on
the same circuit. With a finite number of measurements, you may not converge to
the true expectation value. This is distinct from the notion that running the
same circuit might implement different unitaries if there are uncalibrated
errors. Since we are using expectation values of the quantum circuit to produce
gradient estimates, both of these will be important for a physical
implementation of our algorithm.

\begin{figure}
  \centering
  \includegraphics[width=.85\textwidth]{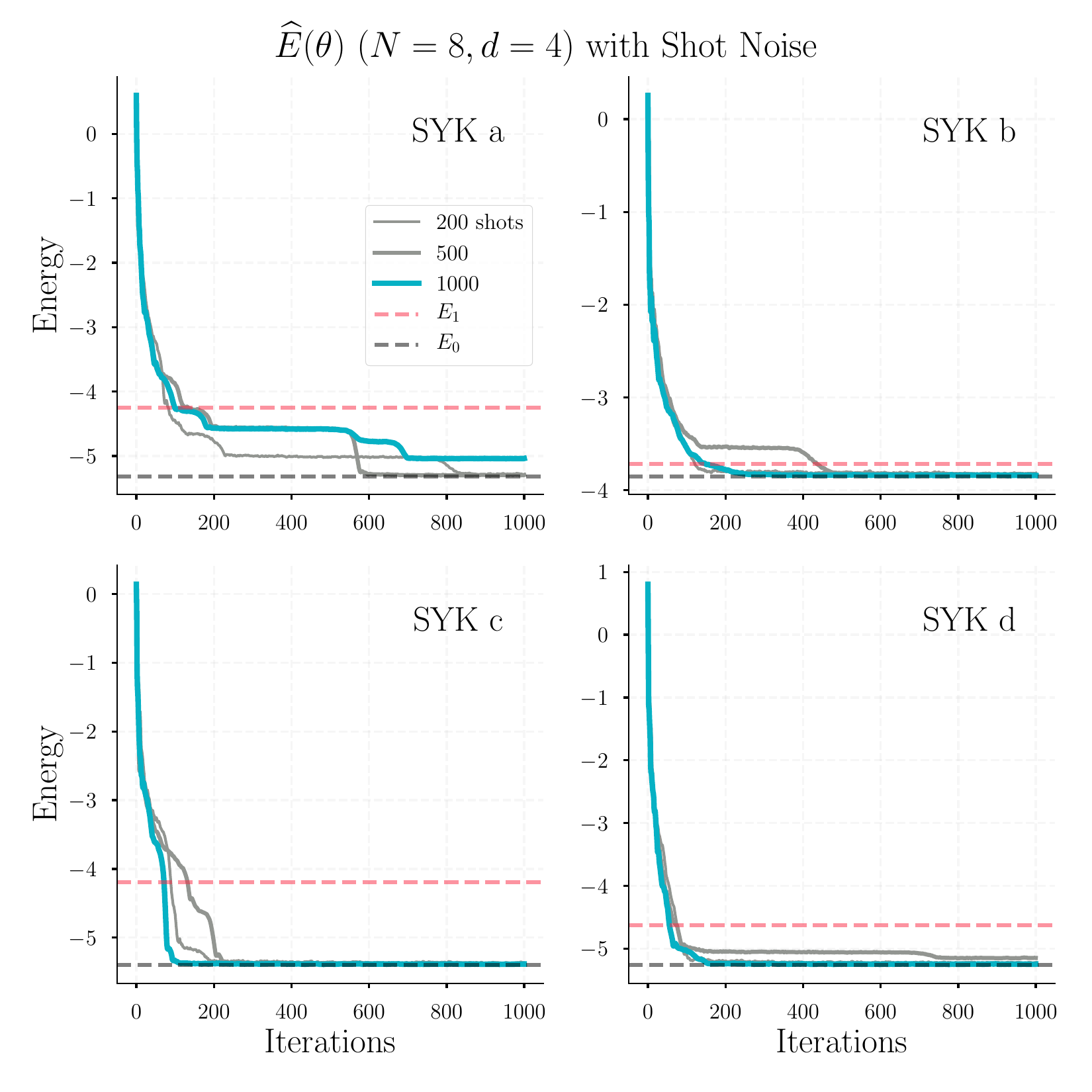}
  \caption{Variational energy achieved when taking shot noise into account.
    First we note that we had to bump up the circuit depth from 3 to 4 to get
    comparable performance. However, this is not a fundamental obstacle to
    achieving energies well below the first excited state. Another interesting
    point is that performance was not monotonic with number of shots. This also
    reflects the difficulty of the optimization landscape.
  }
  \label{fig:training_shots}
\end{figure}

Amazingly, we found that the variational algorithm was able to succeed in the
presence of shot noise. One could worry that the errors in the gradient would render
the algorithm unable to find the ground state energy. Indeed for the same depth
$d=3$, the algorithm did not converge to the ground state energy. However, by
introducing one more layer to our ansatz, we show in Figure.~\ref{fig:training_shots} that
shot noise is not a fundamental obstacle. Interestingly, more shots does not
always guarantee better performance.

\begin{figure}
    \centering
    \includegraphics[width=.85\textwidth]{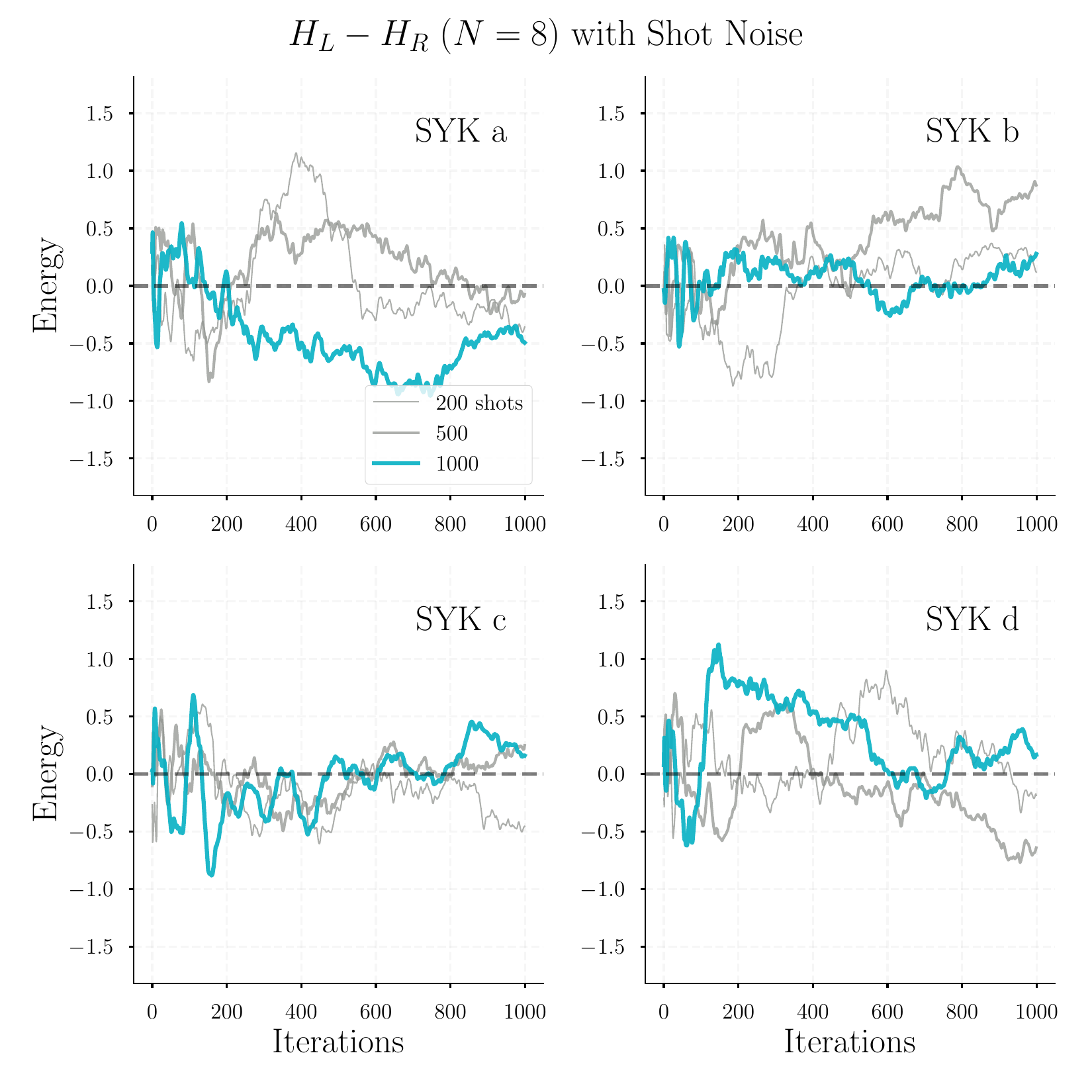}
    \caption{Here we see that in the presence of noise, the parity is violated.
      Even as the ground state energy is approached, the resulting state is not
      necessarily parity invariant. An additional term to the Hamiltonian (the
      objective) that penalizes larger values of $|H_L - H_R|$ could be employed
    to ensure parity.}
    \label{fig:annihilator_shots}
\end{figure}

In Figure.~\ref{fig:annihilator_shots}, we plot the parity violation as a
function of iteration. Generically, there is no reason to expect this to stay at 0
throughout training, which was a particularly special feature of analytic
gradients with the Ising
$R_{XX}$ gate in the previous section. As the ground state energy is approached, this quantity
should tend to 0. The final value we observe is small but nonzero. One could add
another penalty term to ensure parity either in conjunction with the energy or
in an alternating fashion.

\section{Discussion} \label{sec:discussion}
The TFD is a central ingredient for many experiments in recent years with
applications to chaos, scrambling, teleportation and more
\cite{shenker13_black_holes_butter_effec,hosur15_chaos_quant_chann,maldacena15_bound_chaos}. While most works
assume the existence of the preparation, we provide a concrete realization for a
chaotic Hamiltonian, without the need for an external bath.

In this work, we presented a preparation scheme for holographic probes that can
be implemented on a quantum computer \textit{today}. While we are far from the
large $N$ limit, this is an important stepping stone for marching
theory towards experiment and vice versa. We were able to achieve the optimal
energy up to $N=12$ qubits, beyond which exact comparison starts to become
intractable with classical resources.
On the theory side, studying $1/N$ corrections to
the classical limit of Ref.~\cite{maldacena2018eternal} would help us understand how far
we are from a true holographic state with a geometric dual.

Variational computing had been employed previously in constructing the TFD for
the Ising model \cite{Wu_2019,zhu2019generation}, but this approach drastically cuts
down on the classical part of the computation. By getting gradient information
directly from the quantum hardware rather than doing the variational optimization by grid search, one can scale this approach to larger system
sizes. An additional benefit is that the gradient obtained in this way will
factor in the inherent noise due to physical implementation, whereas classical
optimization assumes idealized gates.

We will conclude with some connections and additional directions for future work.

\subsection*{Hardware Implementation}
It would be extremely interesting to prepare the TFD with this variational procedure on a quantum device. It would also enable us to run the traversable wormhole protocol. We make some comments to further close the gap between theory and experiment.

Our choice of PQC was rather uninspiring. We elaborate on why this design choice
is a feature and not a bug. As alluded to in Section~\ref{sec:QC}, gate
compilation is a critical issue for ensuring that a circuit can run on a near
term device.


The beauty in our simple circuit is that there was no optimization of the gates
that indicated this would be optimal, or even remotely good, at preparing the
TFD. Instead, it was built out of gates from a standard gate set that are
perceived to be easy. We expect that hardware
realizations of our procedure would use a variational circuit that is
tailored to what is easily implementable. For fixed gate depth, this prioritizes
the number of ``layers'' rather than worrying about compiling a specific set of
gates. We repeated the experiments with different choice of two qubit entanglers
and achieved similar performance.

Of course, a issue with NISQ devices is the noise. Our simulation
did not include the effects of noise. However, an additional benefit of the
parameter shift rule is that the quantum gradients do not make assumptions on
the form of the unitary being implemented as a classical optimizer would.
Namely, the quantum gradients as measurements yield the gradient for the noisy
gate. This eliminates the need to characterize and calibrate systematic offsets.
Of course, it would be great to verify the gradient procedure working in the
presence of noisy gates on a physical device. An additional hardware
consideration is the tradeoff between circuit depth providing
a more rich ansatz and the additional noise introduced with each layer.

Our approach suggests two ways to implement this in the near term. The first is
to use the simulations to obtain the optimal angles and plug them into a quantum
computer. The more interesting case is to repeat the full procedure and learn on the
quantum device, in regimes we can no longer simulate. Section~\ref{ssec:shots}
provides evidence that shot noise is not a fundamental obstacle to this
procedure working in practice.

\subsection*{Machine Learning}
In this work, we started with a particularly simple form of variational ansatz.
Using an off the shelf gradient descent procedure, we could achieve the ground
state energy of $H_{\text{TFD}}$ for system sizes up to
$N=12$. Both of these seem achievable in the NISQ era, but obviously if their
circuits can be compressed, this would dramatically decrease the time to
successful implementation.
We propose two
engineering approaches to tackling this problem. The first involves
modifying the choice of ansatz with better circuit design, about which little is
known theoretically.\footnote{See \cite{sim19_expres_entan_capab_param_quant} for one
idea to compare the \textit{expressibility} of different quantum circuits.} The second involves trying more sophisticated optimization methods.

On the optimization front, we present two ideas for investigation. The first is
gaining a better understanding of the optimization landscapes of quantum circuits.
Random PQCs have been shown to suffer from the problem of vanishing
gradients \cite{mcclean18_barren_plateaus_quant_neural_networ_train_lands},
rendering the gradient-based navigation much less useful. There have
been recent proposals for training schemes that get around this, such as
training layer by layer \cite{skolik20_layer_learn_quant_neural_networ}.

The second, more ambitious approach would be to use reinforcement learning
methods to automate the circuit design. This has obvious applications to gate
compilation as well, but a successful implementation would circumvent the initial choice of
circuit to compile altogether. Reinforcement learning has been successfully used
in the context of autonomously preparing quantum states of floquet systems \cite{bukov18_reinf_learn_auton_prepar_floquet_engin_states}.

The idea of circuits whose architecture themselves is also variable is something
that has emerged in the quantum chemistry community. See
Ref.~\cite{tang19_qubit_adapt_vqe} and references therein. The algorithm
is driven by constructing the circuit by greedily adding gates from a fixed gate
set that result in a lower energy state.

Whether from a theoretical perspective or a practical implementation view,
there is clear room for improvement when it comes to variational circuit design.

\subsection*{General TFD Preparation}
It would also be interesting to see how to incorporate the more general approach
of TFD preparation as in Ref.~\cite{Cottrell_2019}. Formulating $H_\text{TFD}$ in terms of additional operators would provide an alternate gradient for achieving the TFD state, one whose optimization landscape might be easier to exploit.
An additional theoretical result they show is the existence of a gap in the spectrum of
$H_\text{TFD}$ in their construction even for larger system sizes. As evidenced
in Figure~\ref{fig:spectra}, there was a near degeneracy of the ground state
energy. Having a finite gap for larger system sizes would very likely help our variational
optimization procedure.

An alternative approach to preparing approximate TFD states was discussed in Ref.~\cite{Martyn_2019} in an ansatz called the Product Spectrum Ansatz (PSA).
The PSA is also variational, consisting of a product of mixed states which are then locally entangled with arbitrary 2 qubit gates. A comparison with the ground state method approach in this work is provided in \cite{Martyn_2019}. It would be interesting to see whether one approach would work better in the context of a near term quantum device.

\subsection*{Temperature Dependence}
In this work, we have mostly ignored the thermality of the thermofield double.
It would be very interesting to better characterize the effective temperature of
the system. This could be done analytically, perhaps by a similar analysis as in
Ref.~\cite{maldacena2018eternal}. Other strategies include probing thermal
correlators or comparing numerically to states prepared with imaginary time evolution.

The procedure here was produced for varying $\mu/J = .01, .05, .1,
.5$. For fixed circuit depth, increasing $\mu$ led to a gap in the variational
energy, indicating that it was more difficult to find the ground state. The
subtleties here are related to the fact that we are still very far from the
large $N,\lambda$ limit, so we leave studying the effect of varying interaction strength $\mu$ to future work.

\subsection*{Simulating AdS/CFT}

The role of the SYK model in simulating AdS/CFT was initially proposed in
\cite{garcia-alvarez16_digit_quant_simul_minim_ads}. Towards an experimental
realization of SYK simulation, a proposal using highly controllable ultracold gases was put forth
in \cite{Danshita_2017}. For a review of such holographic quantum matter
proposals, see \cite{franz2018mimicking}. Numerical evolution of the SYK model
was also carried out using classical resources recently in an effort to study
scrambling and chaos \cite{Kobrin20_Many_Body_Chaos_Sachd_Ye_Kitaev_Model}.

With access to the TFD state, this gives a very interesting state to study time
dynamics on a quantum computer. There, we would be interested in simulating time
evolution using digital quantum gates.
One promising avenue for digital Hamiltonian simulation
is qubitization. This approach was taken for the SYK model in
\cite{babbush18_quant_simul_sachd_ye_kitaev}. One could combine the evolution
with this preparation method to carry out the traversable wormhole experiments.

\acknowledgments

It is a pleasure to thank
Unpil Baek,
Ning Bao,
Raphael Bousso,
Nicholas Cinko,
Ben Freivogel,
Tim Hsieh,
Bill Huggins,
Jin-Guo Liu,
Xiu-Zhe Luo,
Hugo Marrochio,
Bryan O'Gorman,
Xiao-Liang Qi,
Pratik Rath,
and
K. Birgitta Whaley
for interesting and useful discussion. Additionally, I thank the group at the NASA Quantum AI Lab for stimulating discussion and hospitality last summer.
I am supported in part by the Berkeley Center for Theoretical
Physics, by the National Science Foundation (award number PHY-1521446), and by
the U.S. Department of Energy under contract DE-AC02-05CH11231 and award DE-SC0019380.
I gratefully acknowledge support by the NSF GRFP under Grant No. DGE 1752814.


\bibliographystyle{JHEP}
\bibliography{references}
\end{document}